\documentclass[12pt,preprint]{aastex}

\newcommand{\Msun}{M$_{\odot}$}
\newcommand{\Rsun}{R$_{\odot}$}
\newcommand{\kms}{km s$^{-1}$}
\newcommand{\etal}{{\it et al.~}}
\newcommand{\ylm}{Y$_l^m$}

\begin{document}

\title{Radial and Nonradial Oscillation Modes in Rapidly Rotating Stars}
\author{C.C. Lovekin \& R.G. Deupree}
\affil{Institute for Computational Astrophysics, Department of Astronomy and Physics, Saint Mary's University}
\email{clovekin@ap.smu.ca}
\keywords{stars: oscillations, stars: rotation}
\begin{abstract}

Radial and nonradial oscillations offer the opportunity to investigate the 
interior properties of stars.  
We use 2D stellar models and a 2D finite difference integration of the 
linearized pulsation equations to calculate non-radial oscillations.  This 
approach allows us to directly calculate the pulsation modes
for a distorted rotating star without treating the rotation as a perturbation.
We are also able to express the finite difference solution in the horizontal
direction as a sum of multiple spherical harmonics for any given mode.  Using 
these methods, we have investigated the effects of increasing rotation and the 
number of spherical harmonics on the calculated eigenfrequencies and eigenfunctions
and compared the results to perturbation theory.  In slowly rotating stars, 
current methods work 
well, and we show that the eigenfunction can be accurately modelled using 
2$^{nd}$ order perturbation theory and a single spherical harmonic.  
We use 10 \Msun models with 
velocities ranging from 0 to 420 \kms\ (0.89 $\Omega_c$) and examine low order
p modes.
We find that one spherical harmonic remains reasonable up to a rotation rate around
300km s$^{-1}$ (0.69 $\Omega_c$) for the radial fundamental mode, but can fail 
at rotation rates as low as 90 \kms  (0.23 
$\Omega_c$) for the 2H mode or $l = 2$ p$_2$ mode, based on the 
eigenfrequencies alone.  Depending on the mode in question, a single spherical 
harmonic may fail at lower rotation rates if the shape of the eigenfunction is 
taken into 
consideration.  Perturbation theory, in contrast, remains valid up to relatively high 
rotation rates for most modes.  We find the lowest failure surface equatorial velocity is 120 \kms
(0.30 $\Omega_c$) for the $l$ = 2 p$_2$ mode, but failure velocities between 
240 and 300 \kms (0.58-0.69 $\Omega_c$)are more typical.

\end{abstract}

\section{Introduction}

Stellar oscillations provide us with a probe of the internal structure of 
stars.  The oscillations depend on the stellar structure, and are modified by
factors such as rotation, magnetic fields and tidal forces.  In theory, if
we have sufficiently accurate parameters for a star, we can produce models 
which will constrain the internal structure.  Unfortunately, due to the
uncertainties on the temperature and luminosity of the star and the large
number of free parameters (mass, rotation rate, age, etc.), this process is 
much more difficult in practise.
Accurate modelling also requires enough observed modes to actually place some 
constraints on the star.  The more modes available, the tighter these 
constraints can be, but we must be sure that all the modes used are real.  
Artificial or extraneous modes can make it impossible to produce a matching
model.
In recent years, the number of stars with multiple modes has 
increased greatly, both thanks to the ground based networks such as STEPHI
\citep{steph} and WET \citep{wet}, as well as space-based observations such as
WIRE \citep{wire} and MOST \citep{most}.  Current and upcoming space missions, 
such as 
Kepler \citep{kepler} and COROT \citep{corot} are expected to further increase 
the number of multiperiodic variables.  Unfortunately, the theory still lags 
behind the observations, particularly for rotating stars.  

The first investigation of non-radial oscillations was undertaken by 
\citet{pek39}.  This paper derived the linearized, 
adiabatic equations for nonradial oscillations of non-rotating stars, and then 
solved the equations for
models of uniform density.  At the time, it was assumed that non-radial modes
would be subject to significant amounts of damping, more so than the purely 
radial modes.  As a result, non-radial oscillations were generally not 
studied extensively.  However, these assumptions do not hold for the low order 
p modes or for all g modes.
Unlike radial oscillations, which are unstable only for $\gamma$ 
$<$ $\frac{4}{3}$, there are some non-radial oscillations of a uniform density
sphere which are unstable
for all values of $\gamma$.  Based on these results, \citet{pek39} concluded 
that non-radial oscillations must be considered.  Using these results, 
\citet{cowl41} calculated the periods of non-radial oscillations for 
non-rotating polytropes.

Before the advent of numerical techniques, these equations had to be solved
using analytical methods.  Much of this work was done by Chandrasekhar, who
explored the variational principle as a method of solving the linear adiabatic
pulsation equations \citep{chand62,chand64}. This method depends on an
arbitrary guess at the form of the eigenfunction, and the resulting eigenvalues
depend on the guess.  Fortunately, even marginal guesses at the eigenfunction
can produce reasonable results for the eigenfrequencies with this method.  This
approach is 
largely unused today, as it has been superseded by computational work using 
more efficient and accurate numerical techniques.

The first direct numerical integration of the linearized equations for 
nonradial oscillations was performed by 
\citet{hrw}.  In this work, they calculated oscillation frequencies for 
non-rotating, polytropic stellar models, for comparison with the earlier 
analytic approaches discussed above.  Although they restricted themselves to
polytropic models, their method can relatively easily be extended to more 
realistic stellar models.

All of these approaches depend on perturbations to a non-rotating (i.e., 
spherical) stellar 
model.  In this case, the calculations are relatively straight forward.  
Rotation, even moderate rotation, can significantly complicate the calculation,
and many attempts have been made over the years to solve the problem with 
varying degrees of success.  These will be discussed in more detail below.  

In spherical stars, the solution to the linear adiabatic pulsation equations 
is separable, and can be written as 
\begin{equation}
\xi_r = X(r)Y^m_l(\theta,\phi)
\end{equation}
The angular variation can be characterized by a single spherical harmonic, 
Y$^m_l$, and both $l$ and $m$ are legitimate quantum numbers.
Once a star becomes distorted, e.g.~through tidal effects or rotation, the 
situation becomes more complex and several problems arise.  The eigenfunction 
can no longer strictly be described by a single spherical harmonic, and thus 
$l$ is no longer a valid
quantum number.  As long as the star remains axisymmetric, $m$ remains 
valid.  As well as changes in the structure of the eigenfunction, the pulsation
frequencies themselves will change.  It is this change in eigenfrequency that
has been of most interest to researchers, particularly as observations continue
to find more and more rotating and pulsating stars, many with multiple 
frequencies.

One of the earlier attempts to solve the linear adiabatic pulsation equations 
for rotating stars was made by \citet{chand62}, who applied the virial
theorem to rotating incompressible fluids.  The variational principle has also
been extended to include slowly rotating stars by \citet{clem64,clem65}. 
Further
attempts at improving the method through a better choice of basis vectors 
have also been made by \citet{clem86}.
Although the variational equations themselves can be applied to a star with an
arbitrary rotation rate ($\Omega$), the method also depends on being able to 
model the structure of the star.  The structure of rotating stars has generally
been modelled as a perturbation to the non-rotating structure.  Because the
structural perturbations are limited to modelling slowly rotating stars, the 
variational method was also limited to slowly rotating stars.

An approach used more frequently now is based on a perturbation approach, as 
developed by \citet{saio81}. In this framework, the rotation is treated as a 
perturbation on the structure of the star. For example, the radial location in 
a rotating model would be written as 
\begin{equation}
	r = a[1 + \epsilon(a,\theta,\phi)]
\end{equation}
The linearized pulsation equations are expanded in a series in powers of the 
rotation rate. The zeroth power merely gives the nonrotating eigenvalues and 
eigenfunctions. Each non-rotating eigenfunction can be written in terms of a single 
spherical harmonic, and the eigenfunction can be characterized by three quantum
numbers relating to the number of radial nodes and the two angular quantum 
numbers, $l$ and m, associated with the spherical harmonics. The first order 
in the expansion in powers of the rotation rate lifts the 2$l$+1 fold 
degeneracy in the eigenvalues, while the eigenfunctions that correspond to 
this order are still characterized by a single spherical harmonic. 

We note that this will not be true in the general set of linearized pulsation 
equations of a rotating star. The coefficients of the perturbations in the 
pulsation equations, composed 
of terms based on the static rotating model, will have latitudinal variations. 
The eigenfunctions will also have a latitudinal variation, so that the 
equations can be expressed as products of spherical harmonics, which in turn 
can be written as sums of spherical harmonics through appropriate recursion 
relations.

In perturbation theory the rotation rate is assumed to be much smaller than the
frequency being calculated. This keeps the rotational perturbation ‘small’ so
that including only the first one or two terms in the power series expansion is
satisfactory. Small is, of course a vague term, and it is not clear how small 
is ‘small’. Based on discussions at the Workshop on the Future of 
Asteroseismology held in Vienna in September 2006, estimates of the limiting 
rotation rate ranged from 50 to 300 \kms. Of course, the limiting surface 
equatorial velocity will be dependent on the mass of the star in question.

Efforts to more accurately include rotation have been developed.  These methods
require 2D calculations, so are more time consuming and complex.  As a result,
previous studies have all faced restrictions and limitations.  For example,
\citet{espin04} calculated the adiabatic oscillations of rapidly rotating stars
with uniform rotation.  To succeed, they applied the Cowling approximation,
neglected the Coriolis force and neglected the Brunt-V\"ais\"al\"a frequency 
in the adiabatic equation. \citet{yoshi01} have modelled quasi-radial modes at 
a range of rotation rates in 
rotating neutron stars using the relativistic Cowling approximation.
Other methods, such as that employed by \citet{lign06} and \citet{reese06} 
have fewer physical restrictions, but have so far been restricted to 
explorations of polytropic models.

The effects discussed in this paper are only expected to matter for stars 
undergoing moderate to rapid rotation.  A recent study of OB stars \citep{daflon07}
found that 50 \% of OB stars have rotation velocities greater than 100 \kms.
At least some of these stars are expected to pulsate.  For example, $\beta$
Cephei-type pulsations have been detected in Spica \citep{sterken86}, which
is also rotating with a vsin$i$ $\sim$ 160 \kms.
For the $\beta$ Cephei stars as a category, the projected rotation velocities
range from 0 to 300 \kms\ \citep{stankov05}.  The average vsin$i$ $\sim$ 100 
\kms, although this could be a selection effect, as the highest amplitude 
pulsators are the more slowly rotating stars.
Another category of pulsating stars, the low amplitude $\delta$ Scuti stars 
(LADS) have been detected with vsin$i$ up to 250-300 \kms  \citep{breger}.
The models we consider in this paper are 10 \Msun\ ZAMS models with solar (Z 
= 0.02) metallicity.  Although $\beta$ Cephei stars have evolved along the 
main sequence, the trends produced by these models should be comparable to 
typical $\beta$ Cephei stars.  One effect which may be important is mode 
bumping, which will appear in real $\beta$ Cephei stars, but does not appear
in our unevolved models.
Our models include uniform rotation at rates from 0 to 0.89 $\Omega_c$.  Our 
method also allows us to consider differential rotation, and this will be 
discussed in a future paper.

\citet{clem98} has developed a finite difference method for 
directly evaluating the eigenfunctions on a 2D grid.  In this paper, we combine
this method with 2D stellar models produced by {\tt ROTORC} 
\citep{bob90,bob95}.  The combination of these two approaches bypasses many of
the restrictions faced by previous approaches.  Our numerical methods and 
models are described in more detail in \S \ref{sec:method}.  We 
investigate the effects of rotation on the calculated eigenfrequencies (\S 
\ref{sec:freq}) and eigenfunctions (\S \ref{sec:func}), with the aim of 
establishing the range of validity of modes calculated with one spherical harmonic.
In \S \ref{sec:pert} we compare our results with those predicted by second 
order perturbation theory.

\section{Method}
\label{sec:method}

Our stellar models are calculated using the 2D stellar evolution code 
{\tt ROTORC} \citep{bob90,bob95},  allowing us to self-consistently model 
the surface and structure of the star for rotation rates from zero up to 
near-critical rotation.  In this paper we focus on uniformly rotating 10 \Msun
ZAMS models with X=0.7, Z=0.02.  We use the OPAL opacities \citet{opalk} and 
equation of state \citet{opaleos} in these calculations.  These models are 
fully 2D, with 10 angular 
zones from pole to equator and 349 radial zones. We have computed a few models 
using 20 angular zones and find differences in the horizontal variation of the 
density to be only about 0.1\%. The pulsation code uses Fourier transform 
interpolation to convert from our angular 
zoning to its own angular zoning, and we feel the {\tt ROTORC} angular zoning 
is not a major source of error in the calculations and use our 10 angular zone 
models in this work.

The location of the surface of the stellar model is found by assuming it lies
on an equipotential surface. The value of the equipotential is determined by 
the value of the total potential in the angular zone which has the largest 
radius (for uniformly rotating models, this is always at the equator). The radial
zone which has this value of the total potential is found at each angular zone 
and the surface boundary conditions applied there. One source of inaccuracy is 
that a radial zone is either completely interior or exterior to the surface, 
so that the surface is defined as the radial zone interface which is closest to
the location of the equipotential. Our rotating models are made by imposing
a surface equatorial velocity and an
internal angular momentum distribution (in this case, uniform rotation) and
allowing the surface to change as needed.  This can lead to small differences
between the imposed (target) surface equatorial velocity and the actual surface
equatorial velocity,
typically less than 2 \kms.  Throughout this paper, we refer to models by
the target surface equatorial velocity.

For our pulsation calculations, we use the non-radial oscillation code (NRO) developed 
by \citet{clem98}.  Instead of expressing the solution as a sum of spherical 
harmonics, the code solves the perturbation equations on a 2D spherical grid.  
In {\tt ROTORC}, the stellar model is defined on a spherical polar grid, with
the stellar surface location being an equipotential surface as discussed above.
NRO tranforms this into a model defined on surfaces of constant density.
The 2D nature of the code allows us to account for the effect of the 
centrifugal distortion, but the Coriolis force is neglected.
The pressure perturbation can be expressed in two ways:
\begin{eqnarray*}
\delta P(r,\theta,\phi;l,m,n) &=& e^{im\phi}\sum_{l=m}^{\infty}a^m_l(r;n,l)P^m_l(cos\theta)\\
{\rm or} \\
 & =& e^{im\phi}\sum_{k=m}^{\infty}A^m_k(r,\theta;n,l)r^k.
\end{eqnarray*}
In this code, the second form of this general equation is used.  Keeping this
general solution in mind, the linear adiabatic pulsation equations can be 
recast using 5 variables, related to the radial and angular velocity 
perturbation, the pressure and gravity perturbations, and the radial derivative
of the gravitational perturbation. These variables are defined as follows:

\begin{eqnarray}
y_1 &\equiv& \frac{\xi_r}{r^{k-1}sin^m\theta}, \nonumber \\ 
y_2 &\equiv& \frac{\xi_{\theta}}{r^{k-1}sin^{m-1}\theta cos\theta},\nonumber \\  
y_3 &\equiv& \frac{\delta p}{r^ksin^m\theta}, \\
y_4 &\equiv& \frac{\delta\phi}{r^ksin^m\theta}, \nonumber\\ 
{\rm and}\;\; y_5 &\equiv& \partial_r y_4 \nonumber
\end{eqnarray}
where $k$ is the radial exponent, $m$ is the azimuthal quantum number, and $k$
= 0 and $m$ = 0 are special cases.  If $k$-1 and $m$-1 are negative, they are 
replaced by 1.  This form of the equations allows the boundary conditions to
be applied while avoiding singularities.  
With these variables, the relevant linearized equations can be expressed in 
the general form
\begin{equation}
\partial_ry_i = f(y_i,\partial_ry_{j\neq i}, \partial_{\theta}y_i)
\label{eqn:gen}
\end{equation}
The full form of the equations and their derivations can be found in 
\citet{clem98}.  

The coefficients of the finite difference expressions of the equations (as 
represented in Eqn.~\ref{eqn:gen}) covering the entire 2D grid can 
be put in a band diagonal matrix.  Each element of this 
band diagonal matrix is itself a matrix, containing the coefficients 
at each zone in the 2D grid.
The solution of the finite difference pulsation equations
proceeds in two steps, from the center outwards and from the surface inwards.  
Each integration also requires an initial guess of the eigenfrequency.  

The inward and outward integrations of the eigenfunctions are required to be 
continuous at some intermediate fitting surface.  
Once all of the coefficients of the equations have been evaluated, a subset of
the matrix, including the fitting surface and the radial zones immediately 
surrounding it can be inverted to solve for the perturbations at
the fitting surface and the radial zone either directly above or below the 
fitting surface.  These values can then be used to step inwards and outwards
 through the 
mesh to solve for the perturbations throughout the rest of the grid.
At some point on the surface,one of the perturbations is forced to be a 
constant (typically, 
$\delta r/r = 1$) to eliminate the trivial solution of all variables being zero
everywhere.  As a result of this, there is one condition that has not been 
used.  This can be used to evaluate a discriminant, which will only be 
satisfied (equal to zero) if an eigenvalue has been located.
Using this method, we can step through eigenfrequency space, solving the 
matrix, evaluating the 
discriminant and looking for zero crossings.  Once a crossing has been located,
various convergence schemes can be used to calculate the exact eigenfrequency.
This method can miss frequencies when two eigenfrequencies are quite
close together, although these can usually be avoided by reducing the frequency
step size.

The code can include up to nine angular zones in the solution for the  
eigenfunctions, performing one radial integration for each 
angle included.  At the end of the calculation, the solution is known at N 
angles, which can subsequently be decomposed into the contributions of 
individual spherical harmonics.  This is done with Fourier transforms, which 
transform the N discrete points into coefficients of the appropriate cosine 
series.  After some
algebraic manipulation, this series is converted into a Legendre series,
which gives us the relative contribution of each \ylm (or Legendre Polynomial
for the case where $m$ = 0).
Because each radial integration contains angular derivatives, also evaluated 
using finite differencing, the resultant coupling among spherical harmonics
arises naturally.  Thus, this method allows us to directly model the coupling
among spherical harmonics in a single pulsation mode for rotating stars in a
natural way.

Because $l$ is not a legitimate quantum number for rotating models, specifying 
$l$ is not necessary. In the pulsation code the input value of $l$ is used to 
specify the parity of the mode, not the exact value of $l$. Based on the parity
of $l$, the code includes 
the first $k$ even or odd basis functions, where $k$ is the input value of the 
number of angular zones to be included. We limit ourselves to small input 
values of $l$ because those are expected to be the most easily observable. We 
also restrict ourselves to axisymmetric modes ($m$=0), although this is not a 
constraint intrinsic to the method. We have also restricted ourselves to modes 
with small radial quantum number ($n$).

Because $l$ is no longer a valid quantum number, we need a new designation for 
mode identification. We have chosen to identify the mode with a quantum number,
$l_0$, which is the value of $l$ of the mode in the nonrotating model to which 
a given mode can be traced back. This tracing back is based on examining both 
the eigenfrequency and the angular shape of the eigenfunction (the modes at 
different radial quantum number are easy to resolve; no mode bumping is
exhibited in these ZAMS models). This is quite easy up to moderate rotation 
rates because one spherical harmonic tends to dominate. This method fails for 
rotation velocities above 420 \kms because no spherical harmonic dominates.  
For rotation velocities above 360 \kms, we find this method becomes somewhat 
uncertain and produces an irregular progression in frequency for some modes.  
We thus consider the pulsation properties for models up to 420 \kms, but regard 
the frequencies above 360 \kms as uncertain.  Although we only consider 
pulsation up to 420 \kms, our static models go up to near critical rotation.

\section{Accuracy of Eigenfrequencies}
\label{sec:freq}

As described in the above section, NRO, combined with 2D structure models 
from {\tt ROTORC}, allows us to calculate the eigenfrequencies for a rotating 
star without making any {\it a priori} assumptions about the structure of the 
star.  
The method of solution of NRO allows for the inclusion of multiple spherical 
harmonics.  As a result, we can calculate eigenfunctions for distorted stars
including the coupling between spherical harmonics.
In contrast most current calculations and observations generally assume that 
pulsation frequencies and observed modes can be characterized by a single 
spherical harmonic. It is therefore of interest to determine at what surface
equatorial 
velocity modes can no longer be adequately described by a single spherical 
harmonic.


One of the issues arising out of the following discussion is where a difference
between two calculated modes becomes significant.  Both ground-based and 
space-based observations continue to improve, as new projects are continuously
launched (figuratively and literally).  As an example, COROT is expected 
to measure frequencies to a precision of less than 0.01$\mu$Hz for the long 
runs, and better than 0.065$\mu$Hz for a faint object during short runs 
\citep{michel07}.  Based on 
these numbers, calculated frequencies do not need to change by much to be
outside the observational uncertainties.  
However, we must ask ourselves whether it is reasonable to expect our models 
to match this accuracy. The linear adiabatic pulsation code uses 10$^{-6}$ as 
the convergence criterion on the discriminant described in \S \ref{sec:method}.
There will be other sources of error on the final eigenfrequency, such as from 
the finite difference representation of the pulsation equations.  Neglecting
these other sources of error, NRO converges modes to 
an accuracy of about 10$^{-6}$, or about 0.001 $\mu$Hz, more than sufficient to 
match the predicted COROT accuracy. However, there are inaccuracies that result
from the finite difference zoning in the static models. When we change the 
surface equatorial velocity from one model to the next, we change the distribution of 
material in the star, although the radial zoning (fractional surface equatorial
radius) remains the same. The changes become larger as the rotation rate 
increases. This is equivalent to changing the radial zoning, which experience 
from the early calculations of linear radial pulsation indicated a sensitivity 
on about the one percent level. We have also fairly dramatically rezoned a 
couple of our models and found that the eigenfrequencies changed on about the 
one percent level, or about 8.5 $\mu$Hz for our models. The higher radial order
p modes are slightly more affected because the outer layers of the model, where
the gradients of model quantities are steeper, play a larger role. Clearly, our
ability to measure
observational frequencies to high precision is irrelevant until models improve
enough to match them.  Until then, for changes induced by rotational effects to
be considered significant, they must be larger than our model uncertainties.

Another uncertainty consideration is the angular resolution of our pulsation 
calculations.  As described above, the number of spherical harmonics used in NRO
determines the number of radial integrations performed.  There are several 
ways we can assess the effects of this changing angular resolution.  Firstly,
we would expect the slowly rotating modes to be relatively unaffected by angular
resolution.  This is indeed what we find.  In the case of slow rotation, the 
coefficients for the higher order spherical harmonics are small, typically not
more than a percent up to 120 \kms.  Over these same rotation ranges, we also
expect the frequency to be relatively unaffected by angular resolution, and
this is indeed what we find.  The frequencies shown in Fig.~\ref{fig:freqchange}
differ by less than a quarter of a percent over this rotation range.

In the majority of our plots, we show our results as a function of surface 
equatorial
velocity, as this is the unit most easily compared to observations.  However,
for comparison with other models, it is more useful to show results as a 
function of angular rotation rate expressed as a fraction of critical rotation 
($\Omega/\Omega_c$).  Critical rotation was calculated using a model rotating
at 575 \kms, with an equatorial radius of 5.792 \Rsun.  This model is quite 
close to critical rotation.  We have summarized the conversion between these 
two frames of reference, as well as some other parameters of our models in Table
\ref{tab:model}. 

\begin{deluxetable}{ccccccc}
\tablecaption{\label{tab:model} Summary of model parameters}
\tablehead{\colhead{target $v_{eq}$ (\kms)} & \colhead{actual v$_{eq}$} & \colhead{$\Omega$ (x10$^{-3}$) (s$^{-1}$)} & \colhead{$\Omega/\Omega_c$} & \colhead{R$_{eq}$ (R$_{\odot}$)} &\colhead{R$_p$/R$_{eq}$} & \colhead{T$_p$/T$_{eq}$} }
\startdata
0   & 0      & 0.0000 & 0.00 & 3.973 & 1.000 & 1.000 \\ \tableline 
10  & 9.97   & 0.0036 & 0.03 & 3.973 & 1.000 & 1.000 \\ \tableline
30  & 29.91  & 0.0108 & 0.08 & 3.976 & 0.999 & 1.001 \\ \tableline
50  & 49.85  & 0.0180 & 0.13 & 3.981 & 0.997 & 1.003 \\ \tableline
90  & 89.72  & 0.0322 & 0.23 & 4.000 & 0.991 & 1.008 \\ \tableline
120 & 119.63 & 0.0428 & 0.30 & 4.021 & 0.986 & 1.013 \\ \tableline
150 & 149.54 & 0.0531 & 0.37 & 4.048 & 0.977 & 1.021 \\ \tableline
180 & 179.45 & 0.0632 & 0.44 & 4.082 & 0.967 & 1.032 \\ \tableline
210 & 209.35 & 0.0729 & 0.51 & 4.125 & 0.953 & 1.051 \\ \tableline
240 & 239.26 & 0.0824 & 0.58 & 4.175 & 0.924 & 1.065 \\ \tableline
270 & 269.17 & 0.0913 & 0.64 & 4.237 & 0.908 & 1.082 \\ \tableline
300 & 299.08 & 0.0998 & 0.69 & 4.307 & 0.887 & 1.100 \\ \tableline
330 & 328.98 & 0.1076 & 0.76 & 4.393 & 0.866 & 1.125 \\ \tableline
360 & 358.89 & 0.1148 & 0.81 & 4.491 & 0.846 & 1.149 \\ \tableline
390 & 388.80 & 0.1215 & 0.85 & 4.600 & 0.821 & 1.173 \\ \tableline
420 & 418.71 & 0.1272 & 0.89 & 4.729 & 0.796 & 1.203 \\ \tableline
\enddata

\end{deluxetable}

\subsection{Frequency Changes}

\begin{figure}
\epsscale{0.8}
\plotone{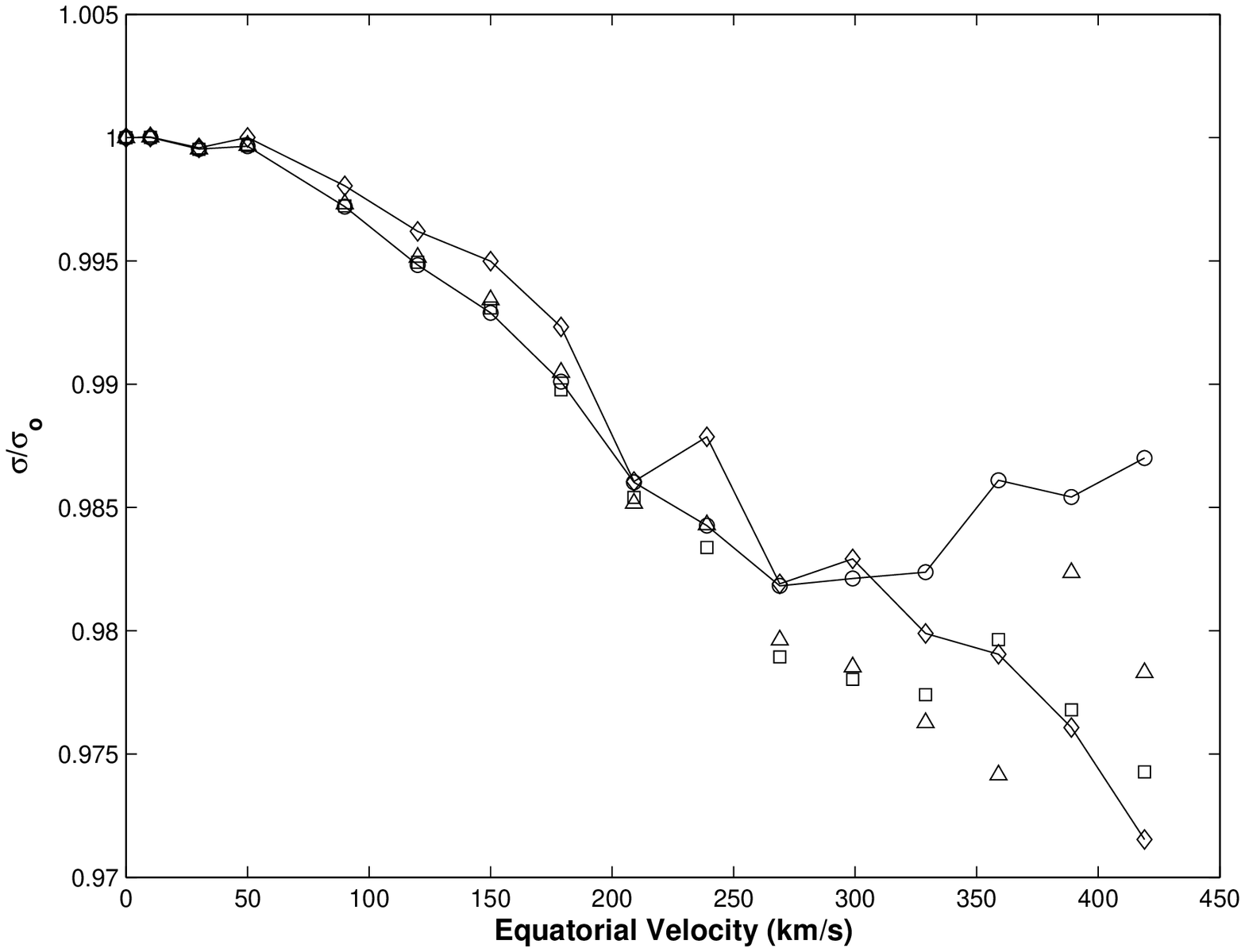}
\plotone{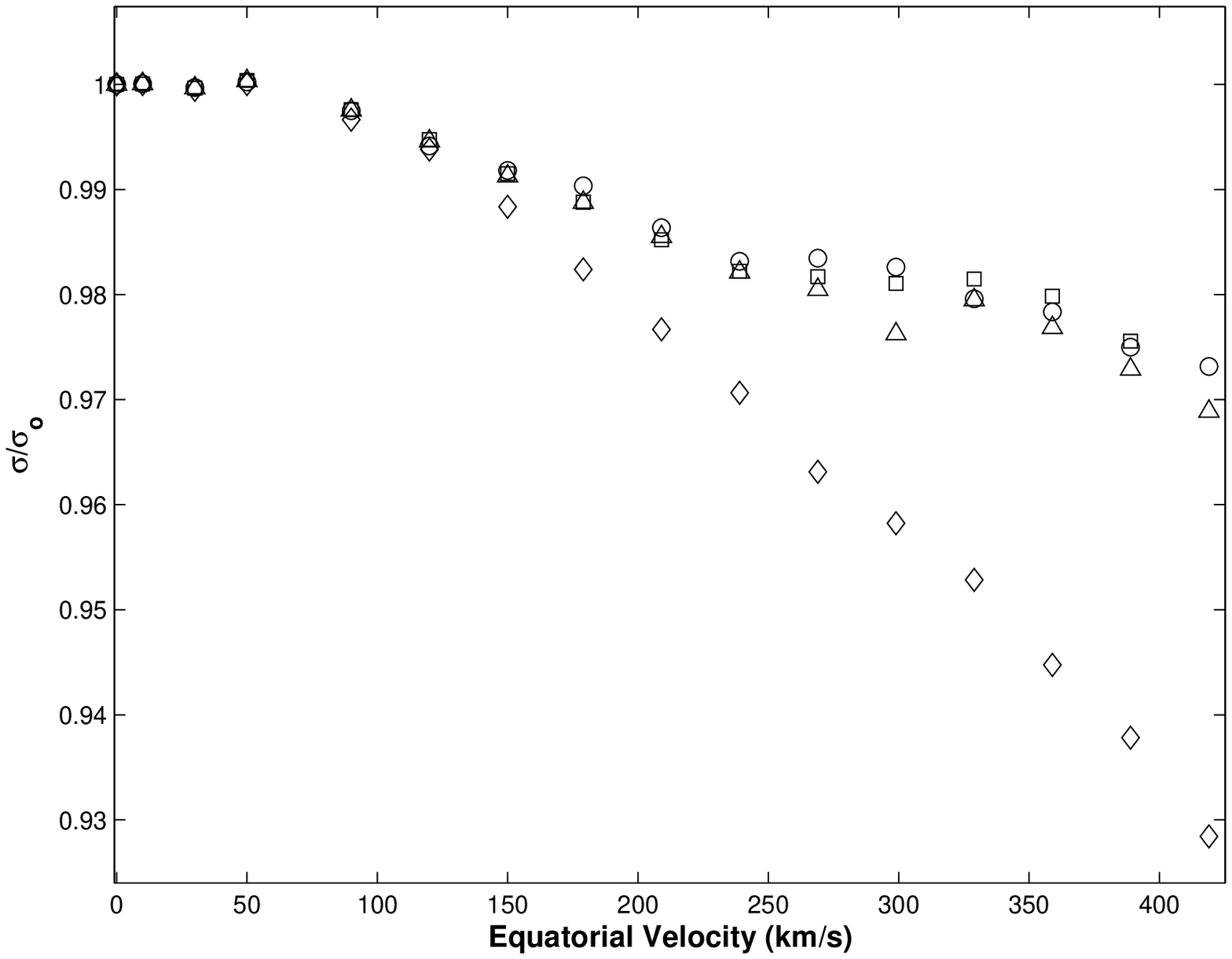}
\caption{\label{fig:freqchange}The frequency changes as a function of surface equatorial velocity for the fundamental mode for $l_0$ = 0 (top) and $l_0$ = 1 (bottom).  Frequencies shown are calculated with ($\Diamond$) - 1 \ylm, $\circ$ - 2 \ylm, ($\Box$) - 3 \ylm, and ($\triangle$) - 6 \ylm.}
\end{figure}

The simplest way to determine where the assumption that a single \ylm can
be used is to compare the frequencies as calculated with different 
numbers of spherical harmonics.  This is illustrated in Fig.~\ref{fig:freqchange}, 
which shows the normalized frequencies for the $l_0$ = 0 and $l_0$ = 1 fundamental
modes, as calculated using 1, 2, 3 and 6 spherical harmonics.  At some cut off 
surface equatorial velocity, the eigenfunctions calculated with only a few spherical harmonics begin
to deviate significantly from those calculated using 6 spherical harmonics.
For the $l_0$ = 0 mode, the frequencies calculated with 1 spherical harmonic are in
reasonably good agreement to quite high velocities, remaining within 
approximately 0.5 \% of the frequencies calculated with more spherical harmonics.  
The $l_0$ = 1 mode as calculated with 1 spherical harmonic rapidly diverges from
the frequencies as calculated with multiple basis functions.  In this case, the
single spherical harmonic frequency reaches a difference of 1\%  at a surface 
equatorial velocity of 210 \kms (0.51 $\Omega_c$).  

Similar results are found for higher order modes.  These results are summarized
in Table \ref{tab:sum}.  To determine the location of the cut off surface 
equatorial velocity,
as described above, we take a difference of 1\% to be significant, as discussed
in \S \ref{sec:freq}.  

Although the periods are expected to change depending on the details of the 
model, period 
differences and ratios are expected to be much more stable.  Hence, in the 
next two sections we will consider the large separation and period ratios of 
our frequency calculations.

\subsection{Large Separations}

\begin{figure}
\plotone{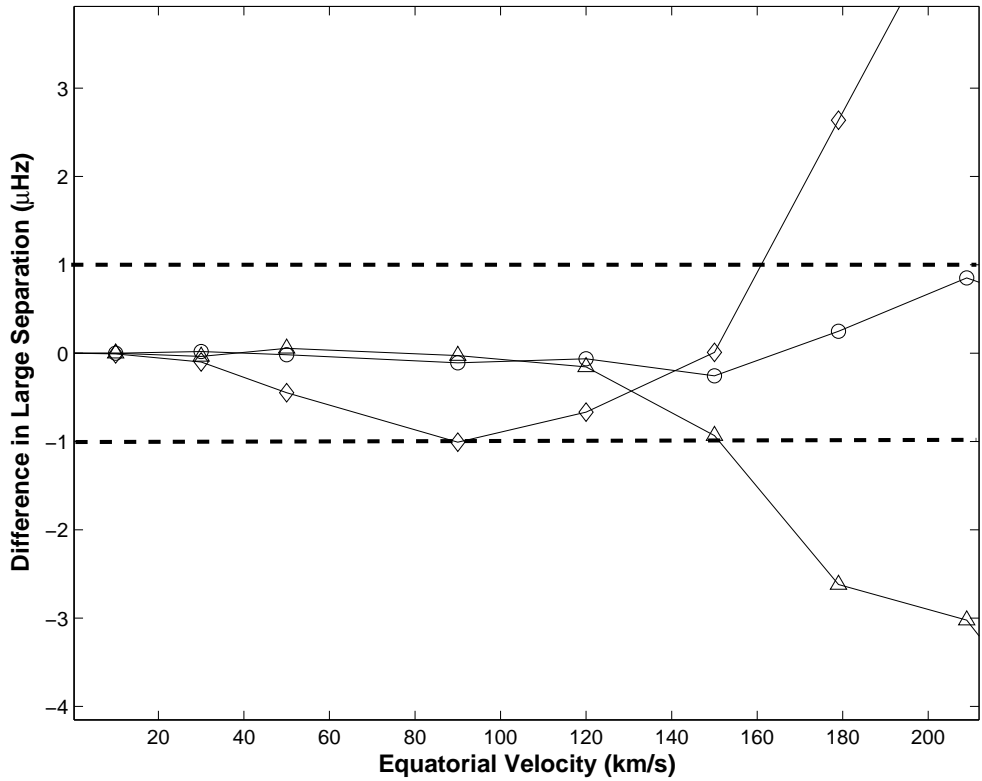}
\caption{\label{fig:largesep}The relative large separation (Eqn.~\ref{eqn:largesep}) as a function of surface equatorial velocity between the $l_0$ = 0 fundamental and first harmonic.  Symbols are as follows: ($\Diamond$) - 1 spherical harmonic, ($\triangle$) - 2 spherical harmonics ($\circ$) - 3 spherical harmonics, all relative to 6 spherical harmonics.  Dashed lines indicate the significance criterion adopted in this work.}
\end{figure}

We have studied the large separation between the $n$ = 0, 1 and 2 modes for $l_0$
= 0-3.  We have calculated the large separations in the usual way
\begin{equation}
\Delta\nu = \nu_{l,n+1} - \nu_{l,n}.
\end{equation}
Before comparing these for the effects of the number of spherical harmonics 
included, we need to account for rotation, which can change
the large separation by changing the model structure.
First, we normalize these large separations with respect to
the non-rotating model
\begin{equation}
D\nu = \Delta\nu(v=0) - \Delta\nu(v).
\end{equation}
We can then use these normalized large separations to look for the effects of 
the number of spherical harmonics included in the calculation ($N$)
\begin{equation}
\label{eqn:largesep}
\mathcal{D}\nu = D\nu_N - D\nu_{N=6}.
\end{equation}
For this calculation, we have assumed that the frequencies calculated with 6
\ylm s are closest to the true pulsation frequencies, so the smaller the 
differences between this and other calculations, the more accurate the smaller
number of spherical harmonics.
This is illustrated in Fig.~\ref{fig:largesep}, which shows the results of
Eqn.~\ref{eqn:largesep} as a function of surface equatorial velocity for the separation between 
the $l_0$ = 0 fundamental and first harmonic.

The uncertainty in the theoretical calculations of large separation is 
inversely proportional to the uncertainty in the radius of the stellar 
model in question.  
Taking the uncertainty in radius to be the size of one radial zone, for our 
models, this is approximately 0.04$\mu$Hz.  Observationally, large 
separations
are well determined for solar type stars, with uncertainties typically less 
than 1$\mu$Hz.  As a conservative estimate, we have chosen 1$\mu$Hz as our
significance criterion, as shown by the dashed lines in 
Fig.~\ref{fig:largesep}.  It should be noted that once the large separations
with 1 and 2 spherical harmonics begin to diverge, they do so quite rapidly, so 
unless the cut off criterion is appreciably smaller ( $\lesssim$ 0.5$\mu$Hz), the
cut off surface equatorial velocity is not an extremely sensitive function of the cut off 
criterion.  The limiting rotational velocities estimated using the large 
separations are summarized in column 4 of Table \ref{tab:sum}.

\section{Accuracy of Eigenfunctions}
\label{sec:func}

So far, the limiting rotation rates entered in Table \ref{tab:sum} have been 
for the $l_0$ = 0 and 1 modes only.  This is a result of the way spherical 
harmonics are included in NRO\@.  To calculate the $l_0$ = 2 mode, for 
example, the code will select even \ylm s starting with $l_0$ = 0, so at least
2 \ylm s are required.  This is true for any mode with $l_0$ $\geq$ 2.  As a 
result, we cannot directly compare eigenfrequencies calculated with several
spherical harmonics to those calculated with a single spherical harmonic.  

We can still compare the eigenfunctions, and in this section this is what we 
will do.  
One of the advantages of including several spherical harmonics is the ability to 
study the effect of rotation not only on the eigenfrequencies, but also on 
the shapes of the eigenfunctions.  For a non-rotating object, regardless of how
many spherical harmonics are included, the eigenfunction remains a pure 
Y$_l^m$,as it should.  
As the rotation rate increases, neighbouring spherical harmonics begin to 
contribute progressively more to the shape of the eigenfunction.  
These effects could be quite important for mode identification, and need to be
considered in rapidly rotating stars.  One technique for mode identification 
uses the pulsation amplitudes in different colors as determined by single 
spherical harmonics.  With
rotation significantly altering the modes by coupling spherical harmonics, 
it could alter these color amplitudes and change the mode identification.
We find that the effects of the coupling can become significant, even at 
very moderate rotation rates.  

\begin{figure}
\begin{center}
\plotone{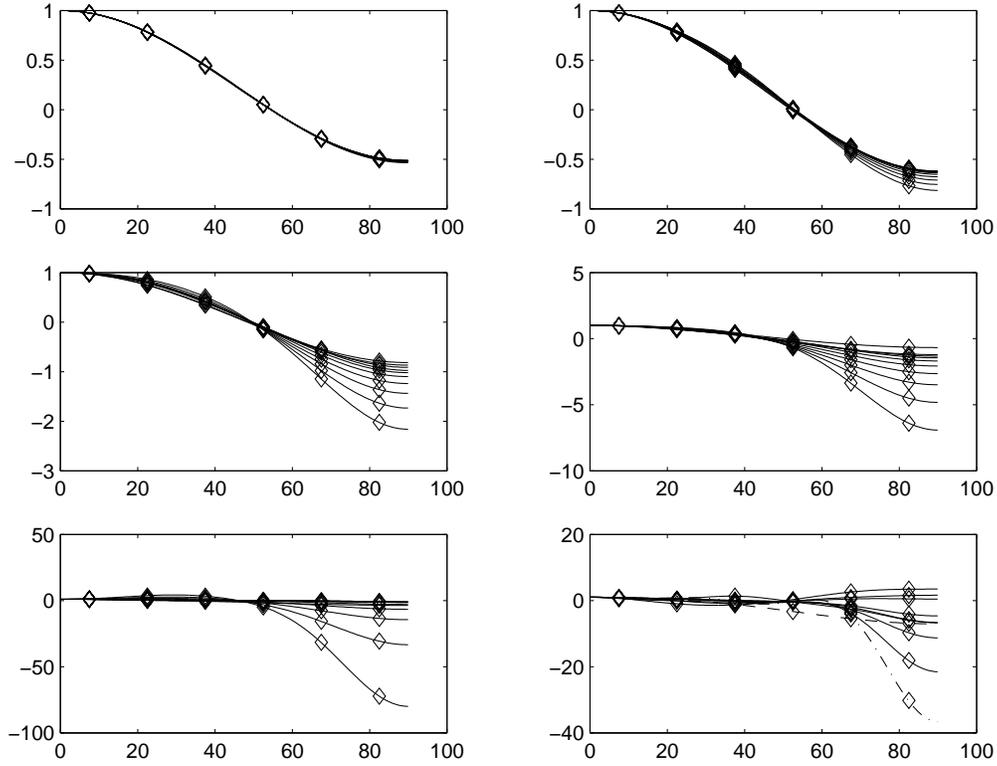}
\caption{\label{fig:depth}Variation in the radial eigenfunction for the $l_0$ mode as a function of colatitude at various depths (fractional surface equatorial radii of approximately 0.1, 0.2, 0.3, 0.4, 0.5, 0.6, 0.7, 0.8, 0.9 and 1.0) for models rotating at 50, 150, 240, 300, 360, and 420 \kms. The convective boundary is located between 0.2 and 0.3 R$_{eq}$.  The variation at each depth is normalized to be unity at the pole for purposes of comparison. The variation is smallest at the center of the star, and increases towards the surface. On the plot of the 420 \kms, the layer closest to the center
is indicated with a dashed line, and the layer closest to the surface is indicated by a dot-dashed line.  In most cases, 420 \kms is the most rapidly rotating model considered, as mode identification becomes difficult.}
\end{center}
\end{figure}

We have used a combination of the value of the eigenfrequency and the angular 
variation of the eigenfunction at the surface to identify the modes as we 
progressed from one rotating model to the next. Of course, with the finite 
difference approach the angular variation of the eigenfunction can vary with 
depth. Fig.~\ref{fig:depth} presents this variation for several rotation rates for
the radial fundamental mode. Each plot contains the variation at several 
different depths. As expected, the variation with depth is small for slowly 
rotating models, and grows as the rotation rate increases. Despite this growth 
in variation, the profile remains recognizably the same until the most rapid 
rotation rate presented. This occurs at a rotation rate at which we are already
beginning to have trouble tracing the modes from one rotation rate to the next 
as we have previously mentioned. 


\begin{figure}
\begin{center}
\plotone{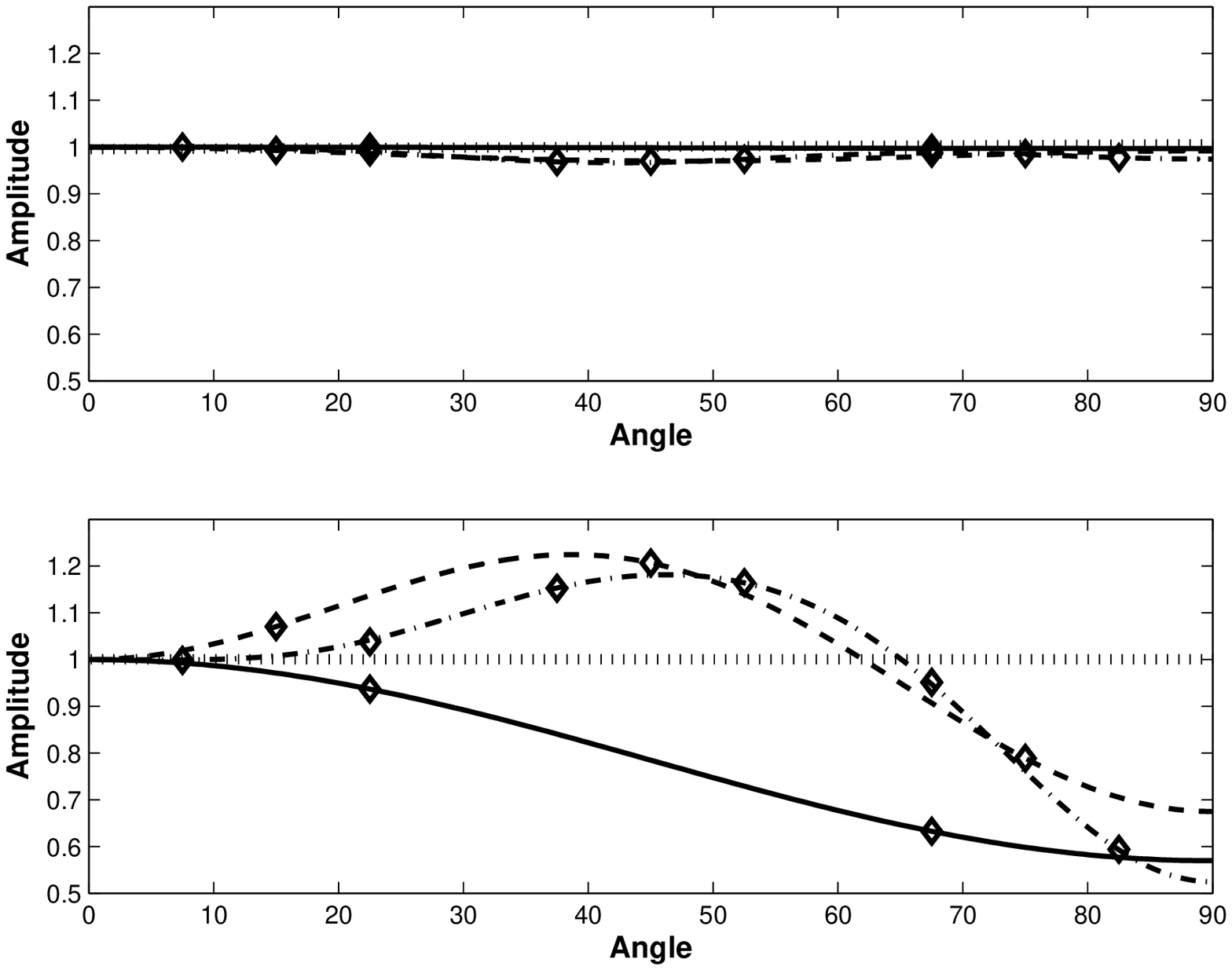}
\caption{\label{fig:l0} 
Angular variation in the radial eigenfunction for the radial fundamental mode of a model rotating at 90 (top) and 270 \kms  (bottom).  On both plots, the shape of the eigenfunction is shown as calculated using 1 (dotted), 2 (solid), 3 (dashed) and 6 (dot-dashed) spherical harmonics.}
\end{center}
\end{figure}

\begin{figure}
\begin{center}
\plotone{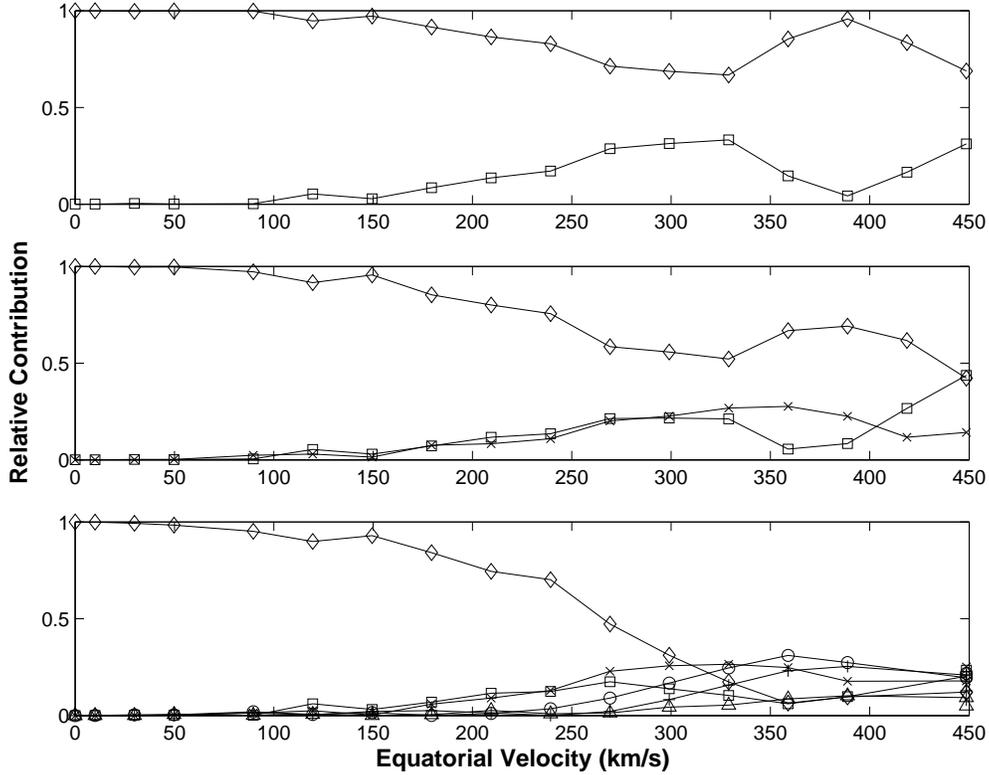}
\caption{\label{fig:trendl0} The relative contribution to the F mode of each spherical harmonic for 2 (top) 3 
(middle) and 6 (bottom) spherical harmonics.  In the top plot, after v $\sim$ 150 \kms, the contribution from $l_0$ = 0
drops below $\sim$ 90\% and we say that you need more spherical harmonics to be able to model the mode.  Symbols are defined as follows: $\Diamond$ - $l=0$, $\Box$ - $l=2$, x - $l=4$, $\circ$ - $l=6$, + - $l=8$, $\triangle$ - $l=10$.}
\end{center}
\end{figure}

Fig.~\ref{fig:l0} shows the angular variation at the surface in 
the radial 
component of the $l_0$ = 0 fundamental mode at 90 and 270 \kms.  At 90 \kms,
the distorting effects of rotation are negligible, although the differences 
are visible.  In contrast,
by 270 \kms  the differences between the numbers of spherical harmonics are quite
significant, and 1 spherical harmonic is clearly not sufficient to model the 
horizontal shape 
of the mode.  In comparison, the eigen\emph{frequencies} were considered to
be accurate using one spherical harmonic up to rotation rates of 300 \kms.  
This highlights the truism that even marginal eigenfunctions can give 
reasonable eigenfrequencies.  
By 270 \kms, the mode no longer looks like an $l = 0 $ mode, nor even an $l=2$,
but is beginning to distinctly show the characteristics of the $l=4$ 
contribution.  These two velocities were chosen based on the relative
contribution of each \ylm, shown for the radial fundamental mode in 
Fig.~\ref{fig:trendl0}.  At 90 \kms, with all three sets of basis 
functions, the $l=0$ component contributes nearly 100\%, while at 270 \kms,
the contribution of the same component drops below 50\% when 6 spherical harmonics
are considered.  

From Fig.~\ref{fig:trendl0}, we can see that with
two and three spherical harmonics, all of the spherical harmonics contribute a 
relatively significant amount by the time the model is rotating at intermediate
speeds.
In contrast, with six spherical harmonics, the contribution from the highest order 
spherical harmonics ($l=10$) remains small out to at least 300 \kms.
Although the contribution starts to become significant at very high rotation
rates (v $\gtrsim$ 350\kms), it still remains a factor of 2-3 lower
than the main contributors.  From this, we have taken the shape of the 
eigenfunction with six spherical harmonics as being the most correct and have 
used it as a basis of comparison.  

\begin{figure}
\begin{center}
\plotone{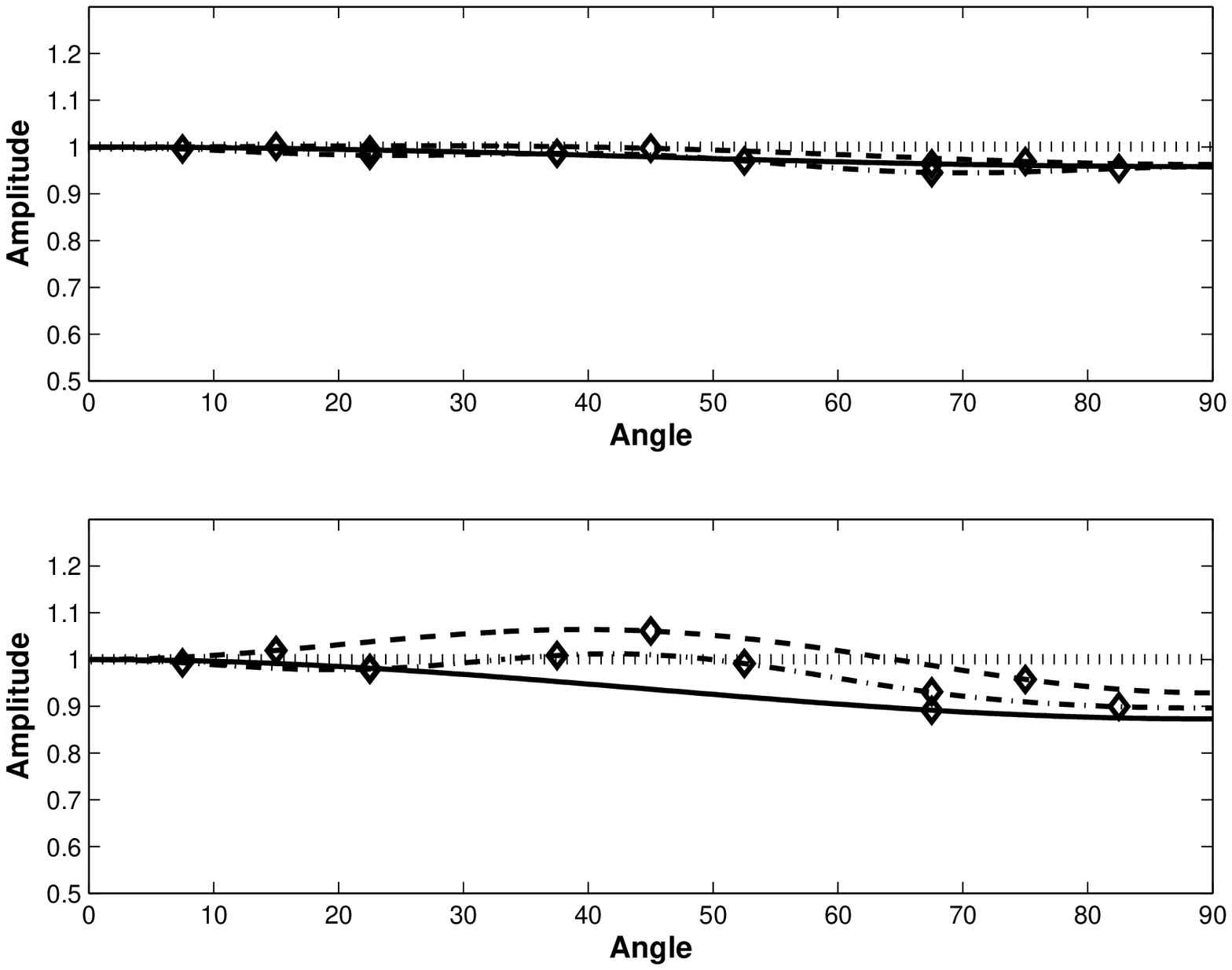}
\caption{\label{fig:cutoff} As for Fig.~\ref{fig:l0}, but for the velocities on either side of the
cut off surface equatorial velocity.  At the lower velocity (150 \kms, top), the shape can be calculated reasonably well using one \ylm, but at the higher velocity (180 \kms, bottom), 2 or more are needed to accurately reproduce the horizontal variation in the eigenfunction.  Symbols are the same as in Fig.~\ref{fig:l0}.}
\end{center}
\end{figure}

Based on the results shown in Fig.~\ref{fig:l0}, we know that one 
spherical harmonic ceases to be sufficient somewhere between 90 and 270 \kms.
From Fig.~\ref{fig:trendl0}, we can see that the relative contribution of the
$Y_0^m$ drops below 90\% at a surface equatorial velocity between 150 and 180 \kms.  
The angular variation of the eigenfunctions for these two velocities are shown 
in Fig.~\ref{fig:cutoff}.
It is at this point that we would say multiple spherical harmonics are required to 
accurately reproduce the shape of the mode (cf. Fig.~\ref{fig:cutoff}).  

\begin{figure}
\begin{center}
\plotone{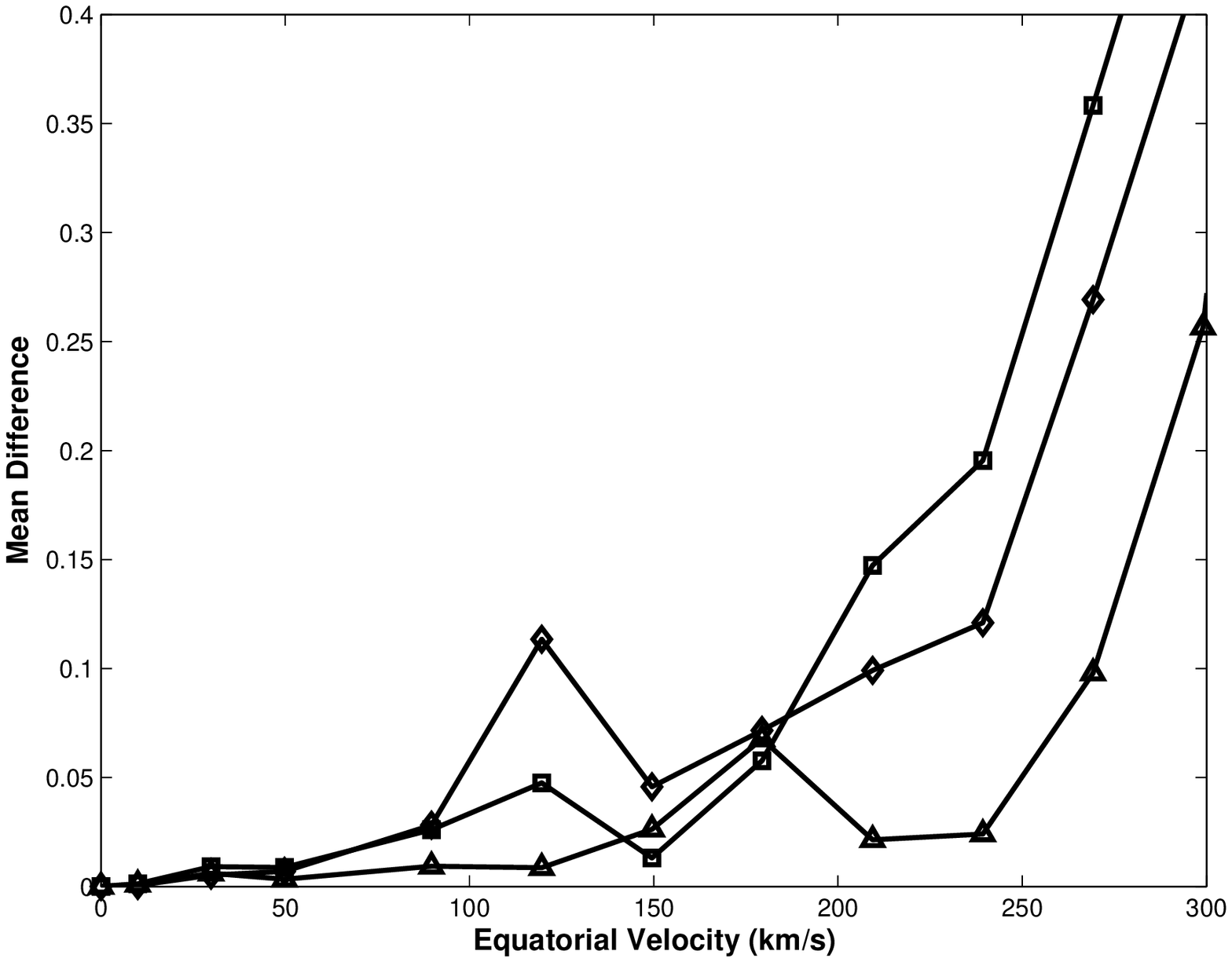}
\caption{\label{fig:diffs} The mean difference between the shape of the radial fundamental eigenfunction with 6 spherical harmonics and a pure $P_0$ mode ($\Diamond$), 2 spherical harmonics ($\Box$) and 3 spherical harmonics ($\triangle$).  Although there is some variation, all three curves show a sharp rise beyond 200 \kms.  See text for the definition of the mean difference.}
\end{center}
\end{figure}

We have developed a quantitative measure of how the shapes of the eigenfunction
differ from that calculated using six spherical harmonics.  This estimate is 
calculated by taking the absolute value of the difference between the 6 basis
function eigenfunction (standard) and one of the other eigenfunctions 
(comparison) at 9 points.  These points are equally spaced across the surface
of the model, with $\theta$ = 10$i$.  The point at $\theta$ = 0 is excluded,
as all the eigenfunctions are normalized to one at this point.  These
differences are then squared and summed.  The
square root of the sum is normalized by the number of points to give a measure
of how different the two curves are:
\begin{equation}
\rm{mean\ difference} = \frac{1}{n}\sqrt{\sum_{i=1}^n (a_i - b_i)^2}.
\end{equation}
This difference as a function of surface equatorial velocity
is shown in Fig.~\ref{fig:diffs}.  The differences between the eigenfunctions
calculated with 1, 2 and 3 spherical harmonics relative to 6 spherical harmonics rises
sharply starting at a surface equatorial velocity of 180 \kms.  Based on this rise
and the eigenfunctions shown in Fig.~\ref{fig:cutoff}, we estimate that when
the mean difference rises above 0.06, more spherical harmonics are needed to 
accurately reproduce the shape of the mode.

For the other modes, the results are qualitatively similar, although the
extent of the differences varies. The results for all four $l_0$ values 
considered in this paper are summarized in Table \ref{tab:sum}.  
Overall, one spherical harmonic remains a good approximation out to at least
90 \kms (0.23 $\Omega_c$).  For some modes, such as the radial fundamental, 
this approximation remains valid to much higher rotation rates (270 \kms, 
0.64 $\Omega_c$).  As both the angular and radial order of the mode increase,
the limiting surface equatorial velocity decreases.  In most cases, we find that the differences
among calculations with different numbers of spherical harmonics grow quickly
as a function of surface equatorial velocity once the differences become sizeable.  We 
can conclude that our results are not 
particularly sensitive to the exact value of the cutoff criterion we have 
chosen, as long as it is not significantly lower than what we have used.  We 
also find that comparing frequencies or frequency differences 
produce approximately the same results.  Based on our results for a 10 \Msun
model, a single 
spherical harmonic is never a good approximation for rotation rates above 0.64 
$\Omega_c$, appears to always be a good approximation for rotation rates below
0.23 $\Omega_c$, and must be used with caution for rotation rates between 
these two values.  Although there may be some mass effects, we do not expect
these results to change significantly for masses close to 10 \Msun.

\begin{deluxetable}{ccccc}
\tablecaption{\label{tab:sum} Summary of velocities at which 1 \ylm fails to accurately reproduce the mode.}
\tablehead{\colhead{$l_0$} & \colhead{$n$} & \colhead{Frequency\tablenotemark{a}} & \colhead{$\mathcal{D}\nu$\tablenotemark{b}} & \colhead{Eigenfunction\tablenotemark{c}} }
\startdata
0 & 0 & $>$360 & -   & 165\\ \tableline 
0 & 1 & 240 & 160 & 60\\ \tableline
0 & 2 & 180 & 24  & 25\\ \tableline
1 & 0 & 210 & -   & 110\\ \tableline
1 & 1 & 210 & 140 & 105\\ \tableline
1 & 2 & 180 & 30  & 85\\ \tableline
2 & 0 & -   & -   & 75\\ \tableline
2 & 1 & -   & -   & 60\\ \tableline
2 & 2 & -   & -   & 45\\ \tableline
3 & 0 & -   & -   & 70\\ \tableline
3 & 1 & -   & -   & 85\\ \tableline
3 & 2 & -   & -   & -\\ \tableline
\enddata
\tablenotetext{a}{~Limiting surface equatorial velocity based on frequency differences larger than 1\%}
\tablenotetext{b}{~Limiting surface equatorial velocity based on difference in the large separation greater than 1$\mu$Hz}
\tablenotetext{c}{~Limiting surface equatorial velocity based on eigenfunctions with mean differences larger than 0.06}
\end{deluxetable}

\section{Comparison with Perturbation theory}
\label{sec:pert}

Second order perturbation theory is routinely used to compute linear pulsation 
modes for rotating stars in which the centrifugal forces are expected to affect
the pulsation frequencies. It has been difficult to comment on when second 
order perturbation theory can be expected to fail because there have been few 
calculations of eigenfrequencies using other methods. Our approach will allow 
placing some limits on the range of applicability of second order perturbation 
theory, but again these limits will be a product of the accuracy obtainable or 
required.

Second order perturbation theory shows that, for axisymmetric modes as we 
consider here, the change in eigenfrequency is a linear function of the square 
of the rotation rate (e.g., Saio 1981). We shall compare our results with this 
linear relation in two separate ways, both of which determine the failure of 
perturbation theory by a deviation from this linear relation. Of course, the 
result will depend on the quantitative value as to when the deviation becomes 
significant, a point we will discuss at the end of this section. We shall use 
the results we feel most accurately reflect the true values of the pulsation 
frequencies, the results with six angular zones in the 2D pulsation grid for 
our comparison of eigenfrequencies.

\begin{figure}
\begin{center}
\plotone{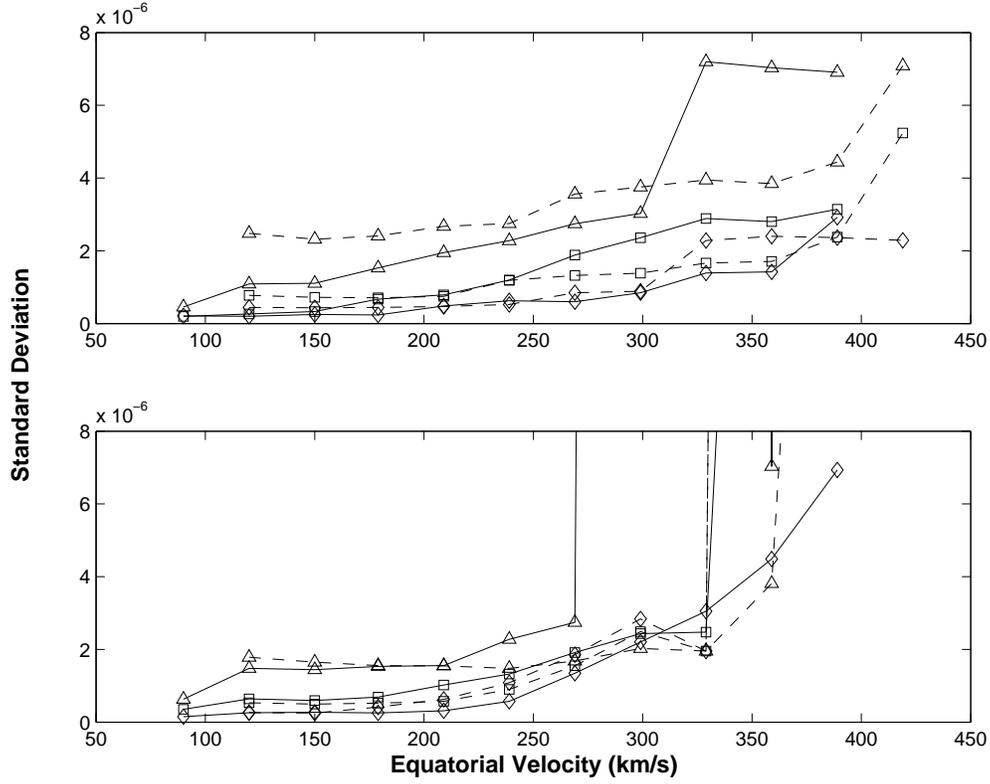}
\caption{\label{fig:freq} Standard deviation from a straight line as more points are included for the $l_0$ = 0 and 1 modes (top) and $l_0$ = 2 and 3 modes (bottom).  We take the cut off standard deviation to be 4x10$^{-6}$.  Symbols are as follows:  $\Diamond$ - fundamental, $\Box$ - first harmonic, $\triangle$ - second harmonic.  Solid lines represent the even modes (0, 2) and dashed lines represent the odd modes (1, 3).}
\end{center}
\end{figure}

The first method starts with the first four models in the rotation sequence 
(surface equatorial rotation velocities from 0 to 90 \kms. We calculate 
the best fit to the linear relationship as given by perturbation theory, and 
the standard deviation. We repeat this exercise, each time adding one more 
model to the analysis, until all rotation velocities are included.  As long as 
the linear relation is satisfied, we expect the standard deviation to be 
approximately constant as we add results for more rapidly rotating models. At 
some point, as the rotation becomes more rapid, the standard deviation will 
become larger and at some threshold value will be declared no longer to be an 
adequate representation of a straight line. Thus second order perturbation 
theory would no longer be considered reliable. We plot this standard deviation 
as a function of the rotation rate of the most rapidly rotating member of each 
sample in Fig.~\ref{fig:freq}.  We somewhat arbitrarily set our threshold at 
4 x 10$^{-6}$ as being a value above the flat region for all modes. The values 
for the limits of applicability of perturbation theory computed by this method 
are listed in the column entitled 'Linear Fit' of Table \ref{tab:pert}.
We have also examined the slope of each linear fit, and as expected, find that
the slope changes gradually where the linear fit is good, and more rapidly as
more points are added.

\begin{figure}
\begin{center}
\plotone{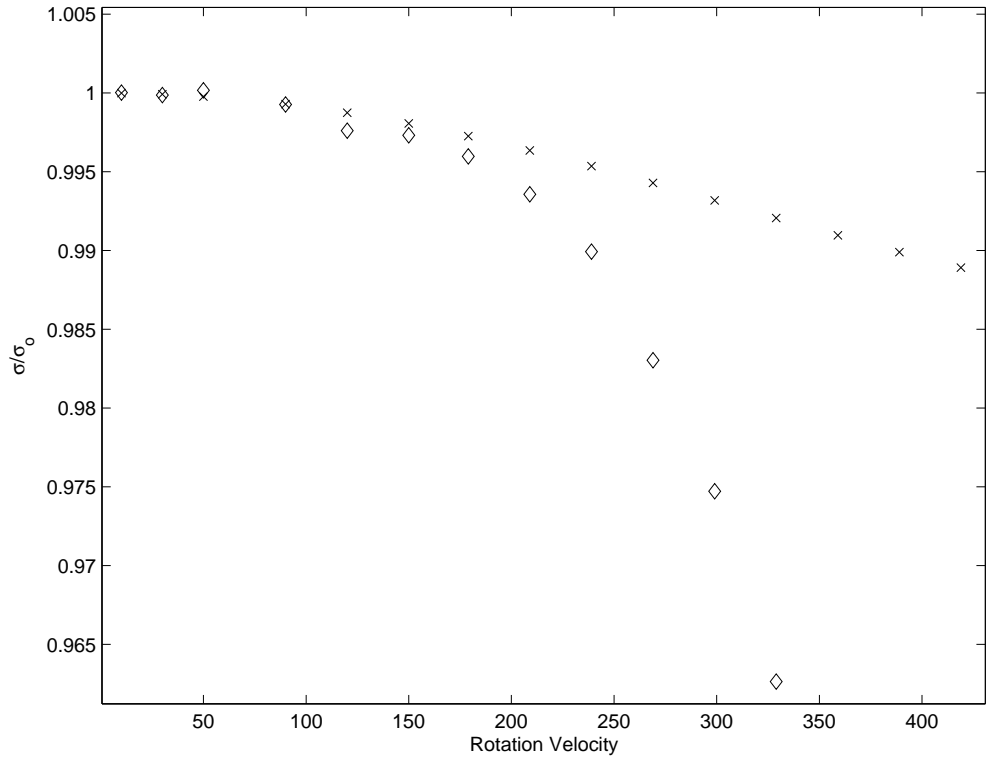}
\caption{\label{fig:pert}Normalized frequencies as calculated with NRO ($\diamond$) and using an estimate of the perturbation theory results (x) for the $l_0$ = 2 f mode.}
\end{center}
\end{figure}

One difficulty with the above approach is that the coefficients of the linear 
fit change as more rapidly rotating models are added. A more constraining 
determination of the threshold of perturbation theory might be obtained by 
using the first few members of the sequence to determine the coefficient of 
the linear fit.  The assumption is that the slope that perturbation theory would
predict is correctly computed using the first few slowly rotating members of 
the sequence. We use the first five members in our rotation sequence to 
calculate this coefficient.  We then use this coefficient to determine 
perturbation theory frequencies at each of our surface equatorial velocities. 
As before, we take the differences between the two methods as significant when
they are larger than 1\%.  
The results for this method are listed in the column entitled
'Coefficient Fit' of Table \ref{tab:pert}. We compare our pulsation 
frequencies with those predicted assuming the coefficient 
computed for the first four members of the sequence is valid at all rotational 
velocities in Fig.~\ref{fig:pert}.

We find the trends for both methods of evaluating the threshold are similar for
the two methods, but that the thresholds computed for the coefficient fit are 
more constrained. This is to be expected because forcing a linear fit to have 
a certain slope is more confining that merely forcing a fit to be linear. It is
interesting that the threshold for perturbation theory occurs at generally 
higher rotation speeds than the threshold for the validity of a single 
spherical harmonic. The extrapolation of the linear fit to higher rotation 
velocities is flatter than our calculation with six angular zones and much 
flatter than our calculation with only one angular zone. 

Our results indicate that perturbation theory is satisfactory to appreciably 
larger rotation velocities than the results of \citet{reese06}, who found that 
third order perturbation theory failed for rotation rates above about 0.2 
$\Omega_c$. Much of this difference arises from the much tighter constraint 
they placed on what difference in eigenvalues is significant. They are able to 
do this because they perform their comparisons using polytropes, which can be 
numerically integrated very accurately, whereas we use finite difference 
techniques to generate our more realistic stellar models. A subsidiary 
consideration is that they can control both the total mass and radius, and thus
can arbitrarily scale from one model to the next, whereas our models include 
the conservation of energy, which removes the radius as an arbitrary parameter.
Also, the surface locations at each angle of our rotating models are quantized;
the surface is regarded to include the full radial zone instead of fractions of
zones. Our errors are in line with variations in eigenvalues computed for 
radial modes at a similar stage of development (e.g., Castor 1971). We believe 
these errors are reasonable at the present time because the deduced properties 
of the stars observed will be inaccurate both from the conversion from observed
parameters to theoretical parameters and from the uncertainties in the effects 
of inclination on the relation between the observed and intrinsic properties. 
The model and parameter inaccuracies will be far greater than the error in the
observed frequencies.  Physical uncertainties, particularly in the internal 
angular momentum distribution, are expected to be greater than or equal to the
uncertainties in an individual model, particularly for the more rapidly 
rotating stars in which we are interested (v > 200\kms).  We believe that 
being able to compute 
the evolution of the rotation law as the star ages may, at this stage, play a 
more important role than increasing the accuracy of the calculations.  Of 
course, we recognize that improvements in accuracy on all fronts are valuable.



\begin{deluxetable}{ccccc}
\tablecaption{\label{tab:pert} Summary of velocities at which perturbation theory fails to accurately reproduce the mode.}
\tablehead{\colhead{$l_0$} & \colhead{$n$} & \colhead{Linear fit} & \colhead{Coefficient fit}& \colhead{1 \ylm (max/min)\tablenotemark{a}} }
\startdata
0 & 0 & -   & 360 & $>$360/160 \\ \tableline 
0 & 1 & -   & 240 & 240/60 \\ \tableline
0 & 2 & 300 & 270 & 180/25 \\ \tableline
1 & 0 & -   & -   & 210/110 \\ \tableline
1 & 1 & $>$360 & 330 & 210/105 \\ \tableline
1 & 2 & 360 & 210 & 180/30 \\ \tableline
2 & 0 & 330 & 240 & 75 \\ \tableline
2 & 1 & 330 & 180 & 60 \\ \tableline
2 & 2 & 270 & 120 & 45 \\ \tableline
3 & 0 & 330 & 210 & 70 \\ \tableline
3 & 1 & 330 & 210 & 85 \\ \tableline
3 & 2 & 360 & 330 & - \\ \tableline
\enddata

\tablenotetext{a}{maximum and minimum rotation speeds at which 1 \ylm is valid, where more than one criterion exists.}
\end{deluxetable}

\section{Conclusions}

In this paper, we have attempted to test the validity of two independent 
assumptions commonly made in calculating stellar oscillation frequencies.  
These are firstly, that 
the non-radial modes can be modelled using a single \ylm, and secondly, that
the modes can be calculated using second order perturbation theory out to some
limiting (highly uncertain) rotation rate.  

We find that when a single spherical harmonic becomes inaccurate is mode 
dependent, with it failing at lower rotation velocities for higher order modes.
The answer is also different depending on what property one examines. A single 
spherical harmonic is sufficient to reproduce frequencies to within 1\% for
rotation velocities up to at least 180 \kms (0.44$\Omega_c$), and for some
low order modes, may even be valid up to 390\kms (0.85$\Omega_c$).  In 
contrast, the angular shapes of the eigenfunctions are extremely sensitive to 
rotation, and the assumption
fails at a maximum surface equatorial velocity of 165 \kms.  In most cases, the 
assumption fails at much lower rotation velocities, typically around 50-75\kms.
Period differences (large separations) are expected to be of most interest,
and these are also found to be sensitive to the order of the mode.  A single
spherical harmonic can accurately predict the difference between the 
fundamental and first harmonic of the $l_0$ = 0 and 1 mode up to velocities of 
around 150\kms (0.37 $\Omega_c$).  The higher order modes are very sensitive to rotation, and 
the assumption fails at velocities of around 25-30 \kms (0.08 $\Omega_c$).
One interesting consequence of the limitations of a single spherical harmonic 
is the impact it may have on mode identification, which is most often based on 
comparing the variation in pulsation amplitude with color with models computed 
assuming a single spherical harmonic (e.g., Heynderickx, Waelkens \& Smeyers 1994). 

We have compared our eigenfrequencies with the relation between eigenfrequency 
and rotation rate predicted by second order perturbation theory. The 
relationship is followed reasonably well for models rotating up to surface 
rotational velocities of about 400 \kms for very low order modes. The relation 
fails at lower rotational velocities (approximately 200 \kms or 
$\Omega/\Omega_c$ = 0.58) for modes with two or three radial nodes. These 
values are dependent on the difference between the two sets of frequencies 
tolerated. In these calculations, the limits are determined by the properties 
of the rotating stellar models rather than the calculations of the 
eigenfunctions.

\acknowledgements

The authors wish to thank M.J. Clement for the use of NRO and for his 
assistance in using and understanding the code.
This research was supported by a Natural Science and Engineering Council of 
Canada (NSERC) Discovery grant and a NSERC graduate scholarship.  Computational
facilities were provided with grants from the Canadian Foundation for 
Innovation and the Nova Scotia Innovation Research Trust.

\end{document}